# Comment on "Is the Hydrated Electron Vertical Detachment Genuinely Bimodal?" J. Phys. Chem. Lett. 2019, 10, 4910-4913

Ruth Signorell

*Department of Chemistry and Applied Biosciences, Laboratory of Physical Chemistry,*

*ETH Zürich, Vladimir-Prelog-Weg 2, CH-8093, Zürich, Switzerland*

Email: rsignorell@ethz.ch

In a recent article [1], David Bartels addresses the issue of the band shape of the genuine binding energy spectrum of the hydrated electron. He essentially claims that the genuine binding energy of the hydrated electron must be Gaussian in shape independent of the photon energy used for ionization (hν), contrary to what was found in ref. [2] for excitations energies in the UV (hν = 3.6-5.8 eV). Further, he alleges that the "bimodal distribution" found for the genuine binding energy in ref. [2] must have resulted from deficiencies in the scattering cross sections used in the scattering calculations. Bartels concludes that these deficiencies arise from an allegedly wrong value of $V_0$ = -1.0eV for the escape barrier used in the original fitting procedure to determine the scattering cross section. In the following, we show that these claims are unfounded and based on partly incorrect assumptions.

The reasoning why the genuine binding energy spectrum should be Gaussian in shape is provided by the author on p.4910, in the middle of the second column in ref. [1]:"If there are a large number of degrees of freedom coupled to the electron, then its ground-state energy and the genuine binding energy function should exhibit a Gaussian distribution by the central limit theorem of statistics". The central limit theorem states that given a sufficiently large set of identically distributed independent random variables, their averages will be distributed normally. This argument cannot be directly applied to the photoelectron spectrum of the hydrated electron. It would paint the picture of the hydrated electron as a structureless entity coupling randomly to a surrounding near-equilibrium bulk liquid. In reality, the hydrated electron consists of the electron itself and the cavity where it resides. From all we know, the vertically ionized state of the hydrated electron, i.e. the empty cavity, is far from any equilibrium structure of the liquid. This could easily give rise to a pronounced vibrational progression in the (genuine) binding energy spectrum of the hydrated electron, especially in the OH-stretch vibration of the water molecules coupling most strongly to it. The central limit theorem describes the inhomogeneous broadening of a transition within a certain limit, but not such homogeneous structures of a transition. Therefore, the central limit theorem cannot predict whether the genuine band shape for excitation in the UV is Gaussian or not.

The recent work of Suzuki and coworkers [3, 4] shows experimentally that the genuine binding energy spectrum for excitations in the XUV (27.9 and 29.45 eV [3]) is indeed Gaussian. The genuine binding energy spectrum, however, depends on both the initial and the final state of the photoionization transition. Therefore, the observation of a Gaussian band shape for XUV excitation cannot predict the band shape for UV excitation. Note that the "retrieval method" in refs. [3,4] does not allow to deduce the genuine band shape either. It just makes the ad-hoc assumption that the band shape is independent of the excitation energy. Thus, it is clear that for UV excitation there is so far no demonstration and no firm argument of a single Gaussian genuine band shape – contrary to ref. [1].

On p.4910, second column, last paragraph in [1], it is stated: "The bimodal distribution is not apparent in the raw data shown in ref. [2] and only appears after the analysis with low-energy scattering cross sections from ref. [5]." This mere claim remains unsubstantiated in ref. [1] and is in fact incorrect (see below and Fig. 1). Correct is that the asymmetric ("bimodal") genuine band shape we obtained in [2] directly reflects the asymmetric and unexpectedly broad band shapes of the experimental ("raw") data obtained at the higher UV photon energies (see e.g. Figs. 2 and 3 in [2]), and does not "only appear after the analysis with low-energy scattering cross sections from ref. [2]". The asymmetric and broad band shape of the experimental spectra at the higher energies are not consistent with a single Gaussian genuine binding energy (Fig. 1). The "bimodal" genuine band shape represents the properties of the experimental spectra and is not an artifact of deficient scattering cross sections as ref. [1] would have it.

The statement on p.4911, left column, before the last paragraph in [1]: "It is clear that the value of $V_0 = -1.0eV$ used in the fitting of Luckhaus et al. is inconsistent with their own data." is wrong. It is based on the assumption that $V_0$ should be given by the difference between the photoconductivity threshold and the onset (threshold) of the binding energy spectrum. This is physically not correct because the electron escape occurs on a timescale much faster that the relaxation times in liquid water. That means that it occurs in the sudden limit and not in the adiabatic limit. Therefore, the only physically plausible estimate of $V_0$ must refer to the vertical and not the threshold binding energy. It is thus clear that Michaud et al. [6] based their estimate of $V_0 = -1.0eV$ on a physically sound argument. There is no inconsistency between this estimate and the genuine binding energy spectrum we derive in [2].

In the last paragraph on p.4911, left column in [1], the author tries to backup his claim of inconsistency with the value of $V_0$ "between 0 and -0.12eV" derived by Coe et al. [7]. It has already been shown in the literature (e.g. refs. [8, 9]) that the result of Coe et al. to have determined a lower limit for $V_0 = -0.12eV$ is not correct. As explained in those two references, the derivation by Coe et al. contained a sign error in the reorganization energy of water. With the correct sign of the reorganization energy, the only conclusion Coe et al. can draw from their data is that $V_0 \geq -1.72eV$. This result is obviously perfectly consistent with the estimate by Michaud et al. [6], which we adopted in our work [2]. Contrary to what the author of ref. [1] states in the last paragraph on p.4911 left column and continued in the right column, the correct lower bound of $V_0 = -1.72eV$ is fully consistent with the value we used ($V_0 = -1.0eV$) for the determination of the scattering cross sections [2, 5].

In the last paragraph of ref. [1] the author suggests that we should refit our scattering cross sections "with a more realistic choice of $V_0$" (i. e. with $V_0 \sim -0.1eV$), and predicts that the refitted cross sections would result in a genuine binding energy spectrum with Gaussian shape. Fig. 1 illustrates that this is not the case. A single Gaussian genuine spectrum with cross sections refitted for $V_0 = -0.1$ eV (red line) does not reproduce the experimental spectrum (black line). It is precisely the signal observed at high electron binding energies (eBE) that a single Gaussian fails to reproduce. Virtually the same results are obtain with $V_0 = -1.0$ eV and only a single Gaussian (blue line). Therefore, the prediction in the last sentence in the conclusion of ref. [1] on p.4911: "When these refitted cross sections are eventually applied to the experiments in 1, we predict that a genuine eBE(g) function will emerge that is very nearly Gaussian…" is not correct either.

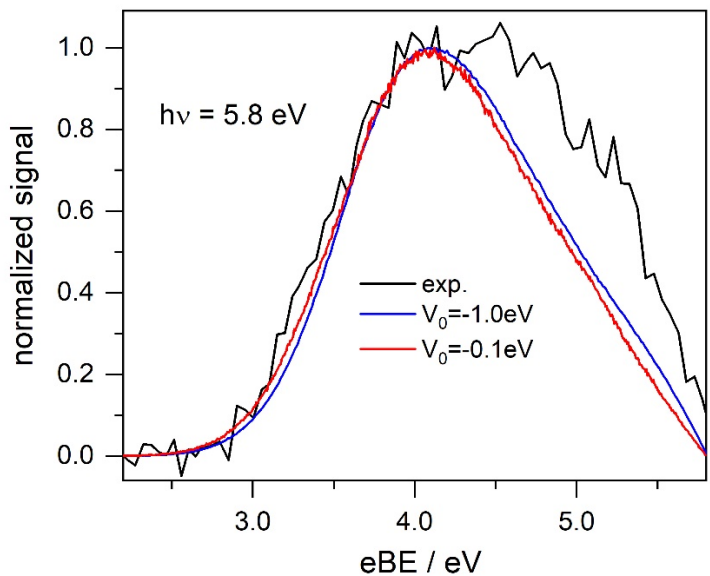

**Fig. 1:** Black line: Experimental binding energy spectrum of the solvated electron in liquid water recorded at a photon energy $h\nu = 5.8$ eV [2]. Red line: Binding energy spectrum calculated with cross sections refitted for an escape barrier $V_0 = -0.1$ eV and a single Gaussian genuine band shape. Blue line: Binding energy spectrum calculated with the original cross sections and escape barrier of $V_0 = -1.0$ eV, and a single Gaussian genuine band shape.

In summary, the asymmetric ("bimodal") genuine band shape in the UV we determined in ref. [2] is not an artefact of the scattering cross sections based on current knowledge. Currently, there is no firm evidence for the genuine binding energy spectrum of the hydrated electron to be Gaussian in shape or to be independent of the photon energy used for excitation as assumed in ref. [1]. There are two more possible explanations for the observed asymmetric ("bimodal") genuine band shape for UV excitation to be clarified in the future: (i) Either the genuine band shape in the UV is indeed bimodal and thus different from that in the XUV [3], or (ii) there are measurement artefacts in the experimental UV spectra used in in ref. [2] that cause the bimodal genuine distribution. Currently there is not sufficient evidence to decide whether (i) or (ii) or both are true. Note that new measurements in the UV would require photon energies clearly above 4.6eV to be able to probe a possible bimodal component in the genuine spectrum. We agree with the author of ref. [1] that the genuine band shape of the hydrated electron in the UV is an exciting scientific question to be clarified in the future.